\documentclass[a4paper, 12pt]{amsart}

\usepackage{amsmath}
\usepackage{amssymb}
\usepackage{amsthm}
\usepackage{graphics}
\usepackage[usenames,dvipsnames]{xcolor}
\usepackage{url}

\definecolor{DeepBlue}{rgb}{0,0,.7}
\usepackage[dvipdfm, setpagesize=false]{hyperref}
\hypersetup{colorlinks=true, citecolor=Sepia, linkcolor=DeepBlue, urlcolor=DeepBlue, pdfborder= 0 0 0}
\hypersetup{setpagesize=false}
\usepackage[all]{hypcap}

\newcommand{\F}{\mathcal{F}}
\newcommand{\s}{\mathrm{sgrad\ }}
\newcommand{\D}{\mathrm{d}}
\newcommand{\Lin}{\mathrm{Lin}\ }
\newcommand{\Ker}{\ker}
\newcommand{\rk}{\mathrm{rk}\ }
\newcommand{\h}{f}
\renewcommand{\sp}{sp}
\newcommand{\R}{\mathbb{R}}
\newcommand{\C}{\mathbb{C}}
\newcommand{\Z}{\mathbb{Z}}
\newcommand{\dd}{\partial/\partial}
\newcommand{\bd}{\partial}

\newcommand{\FC}{C_2}

\newcommand{\prop}{Proposition}

\newtheorem{Th}{Theorem}[section]
\newtheorem{Lemma}[Th]{Lemma}
\newtheorem{Proposition}[Th]{Proposition}
\newtheorem{Corollary}[Th]{Corollary}
\theoremstyle{definition}
\newtheorem{remark}{Remark}[section]
\newtheorem{definition}{Definition}[section]
\newtheorem{condition}{Condition}[section]

\begin{document}

\title{
Singularities of integrable Hamiltonian systems:
a criterion for non-degeneracy,
with an application to the Manakov top
}
\author{Dmitry Tonkonog}
\thanks{The author was partially supported 
by Federal Target Program grant 02.740.11.5213 ``Bi-Hamiltonian
structures and singularities of integrable systems''
(research group of Prof.~A.V.~Bolsinov) in 2010/2011,
by Dobrushin Scholarship at the Independent
University of Moscow and by Special Scholarship of the Government of Russia
in 2011.}
\address{Department of Differential Geometry and Applications, Faculty of Mechanics and Mathematics,
Moscow State University, Moscow 119991, Russia.}

\email{dtonkonog@gmail.com}


\begin{abstract}

Let $(M,\omega)$ be a symplectic
$2n$-manifold and $h_1,\ldots,h_n$
be functionally independent commuting functions on $M$.
We present a geometric criterion for
a singular point $P\in M$
(i.e.\ such that $\{\D h_i(P)\}_{i=1}^n$ are linearly dependent)
to be
non-degenerate in the sense of Eliasson--Vey.

The criterion is applied 
to find non-degenerate  
singularities in the Manakov top system 
(aka the 4-dimensional rigid body).
Then we apply Fomenko's theory
to study the neighborhood $U$ of the singular
Liouville fiber containing  
saddle-saddle singularities of the Manakov top.
Namely,
we describe the singular Liouville foliation
on $U$
and 
the `Bohr-Sommerfeld' lattices
on the momentum map image of $U$.
A relation with the quantum Manakov top studied by
Sinitsyn and Zhilinskii (SIGMA 3 2007, arXiv:math-ph/0703045)
is discussed.
\end{abstract}

\maketitle

\setcounter{tocdepth}{1}

\tableofcontents

\section{Introduction and a criterion for non-degeneracy}
\label{SecInt}
This paper is on singularities of
Liouville integrable
Hamiltonian systems.

First we briefly present basic definitions used in the paper.
A {\it Liouville integrable Hamiltonian system (IHS)}
$(M,\omega,h_1,\ldots,h_n)$
is a symplectic
$2n$-manifold $(M,\omega)$ with
functionally independent commuting functions
$h_1,\ldots,h_n:M\to \R$
traditionally called {\it integrals}.
(For our purposes it is not important which of them is
the actual Hamiltonian and which are additional integrals.)
For a function $g$ on $M$,
its Hamiltonian vector field
is denoted by $\s g$.
The
{\it momentum map} $\F:M\to \R^n$ is
given by $\F(x):=(h_1(x),\ldots,h_n(x))$.
Level sets of $\F$ 
(that is, commom level sets of $h_1,\ldots,h_n$)
are called {\it Liouville fibers}.
A point $x\in M$ is called
a {\it singular (critical) point of rank $r$},
$0\le r<n$,
if $\rk \D\F(x)=r$.
The $\F$-image of all singular points
is called the {\it bifurcation diagram}.
For singular points, there is a natural notion of
non-degeneracy \cite{Eli90}, \cite[Definition~1.23]{BF04}.
Now we recall this definition for zero-rank
critical points (the general definition is given below),
and then 
describe the structure and main results of the paper.

\begin{definition}
\label{defnondeg1}
Let $(M,\omega,h_1,\ldots,h_n)$
be an IHS and $P\in M$ be a zero-rank singular point,
i.e.\ $\D h_i(P)=0$ for each $i$.
The point $P\in M$ is called
{\it non-degenerate} if the commutative
subalgebra $K$ of $sp(2n,\R)$ generated by
linear parts of Hamiltonian vector fields $\s h_1,\ldots,\s h_n$
at point $P$
\footnote{
Equivalently, $K$ is generated by linear
operators $\{\omega^{-1}\D^2h_i(P)\}_{i=1}^n$.
The commutativity of $K$ is implied by the fact
that $h_i$ commute.
}
is a Cartan subalgebra of $sp(2n,\R)$.
\end{definition}
\nopagebreak[4]
{\it Structure of the paper.}
In this section we present Theorem~\ref{T1} (main result),
which is a geometric
criterion for non-degeneracy of zero-rank
singularities {\it of elliptic-hyperbolic type} 
(see Remark~\ref{RemConverse}), 
and Theorem~\ref{T2} extending Theorem~\ref{T1} on
singularities of arbitrary rank.
We prove both theorems in \S\ref{SecProofCrit}.
In \S\ref{SecManakov} we study the Manakov top
system, aka the
4-dimensional rigid body.
Namely, we apply Theorem~\ref{T1} to find
non-degenerate singularities of the Manakov top
(\prop~\ref{ThNondegen})
in terms of the bifurcation diagram.
After that we study the 4-dimensional
neighborhood $U$ of the singular Liouville fiber
containing
{\it saddle-saddle} 
(see Definition~\ref{defwilliamsontype})
singularities
of the Manakov top.
The proved non-degeneracy  
allows us to
describe in \prop~\ref{ThSemiprod} the singular {\it Liouville foliation}
(i.e.~foliation on level sets of $\F$) on $U$
very easily, just by finding the correct
alternative from
the complete list of singularities 
obtained by Fomenko and his collaborators
\cite[Tables 9.1 and 9.3]{BF04}.
Purely topological
\prop~\ref{ThSemiprod}
has an interesting application,
\prop~\ref{ThLattice}.
It describes the {\it `Bohr-Sommerfeld lattice'} of the Manakov top
which we define as 
the momentum map image of those Liouville tori in $U$
on which the action variables take values in $2\pi h\Z$,
(ignoring for simplicity any Maslov-type correction).
Proofs of statements from \S\ref{SecManakov}
are given in \S\ref{SecProofManakov}.

\smallskip
{\it Relations with other results.}
Singularities of the Manakov top were
previously studied in \cite{Osh87, Osh91, Au02, FM03, SiZh07, BCRT, BC},
see \S\ref{SecManakov} for details.
In particular, the recent paper \cite{BCRT}
obtains a complete comprehensive description of 
non-degenerate singularities of the Manakov top,
from which
\prop~\ref{ThNondegen}
could be deduced.
However, the proofs in \cite{BCRT} involve rather long computation;
the proof of \prop~\ref{ThNondegen}
using Theorem~\ref{T1} is considerably shorter.

The problem to describe the structure 
of saddle-saddle
singularities of the Manakov top
was raised
in \cite{SiZh07}
during analysis of the quantum Manakov top.
In this paper, Sinitsyn and Zhilinskii 
numerically calculated and visualized 
\cite[figures 1 and 13]{SiZh07}
the joint spectrum lattice of two operators
corrresponding to the quantum Manakov top.
This lattice is very similar to the 
`Bohr-Sommerfeld lattice' described in
Proposition~\ref{ThLattice}.
We discuss this
in the end of \S\ref{SecManakov}.
The two lattices are available for
comparison on fig.~\ref{figManakovLattice}.

\smallskip
Now we briefly discuss the notion of non-degeneracy
to motivate Theorem~\ref{T1}.

In general, non-degenerate singularities
are important because they are generic and
because 
the local structure of integrable systems
in their neighborhood 
is well understood, see Theorem~\ref{ThNF}.
Global structure of non-degenerate singularities
(i.e. structure of neighborhoods of whole Liouville fibers
containing non-degenerate singularities)
was studied by Fomenko and his school, 
as well as by others;
see survey \cite{BO06}, book \cite{BF04}
and papers \cite{NTZ95, NTZ96, LeUm98, Ler00, Osh10}.
The following is the fundamental fact about 
non-degenerate singularities, cf. Remark~\ref{remsmooth}.


\begin{Th}[on Normal Form]
\label{ThNF}
\cite{Ru64, Vey78, Ito89}.
Let $P\in M$ be a non-degenerate zero-rank singular point 
of an analytic IHS $(M,\omega,h_1,\ldots,h_n)$.
Then there exist a local 
system of coordinates $(p_1,\ldots,p_n,q_1\ldots,q_n)$ at point $P$
and nonnegative integers $m_1,m_2,m_3$
with $m_1+m_2+2m_3=2n$
such that $\omega=\sum_{i=1}^n dp_i\wedge dq_i$ and
for each $i=1,\ldots,n$ we get 
$h_i=h_i(G_1,\ldots,G_n)$ where
\begin{eqnarray*}
G_j =p_j^2 + q_j^2  
&
\mbox{(elliptic type)}
&
j=1,\ldots,m_1
\\
G_j = p_jq_j
&
\mbox{(hyperbolic type)}
&
j=m_1+1,\ldots,m_2
\\
\left.   
\begin{array}{ll} G_j &= p_j q_{j+1}- p_{j+1} q_j \\
                  G_{j+1} &= p_j q_j + q_{j+1} p_{j+1}
\end{array} 
\right\} 
&
\mbox{(focus-focus type)}
&
\begin{array}{r}
j=m_1+m_2+1,\ m_1+m_2+3,\ldots\\
\ldots,m_1+m_2+2m_3-1.
\end{array}
\end{eqnarray*}
\end{Th}

\begin{definition}
\label{defwilliamsontype}
The triple $(m_1,m_2,m_3)$
is called the {\it Williamson type}
of $K$, cf. \cite{Wi36}. In the case of two degrees of freedom
($n=2$)
these types are also called: 
{\it center-center} $(2,0,0)$,
{\it center-saddle} $(1,1,0)$,
{\it saddle-saddle} $(0,2,0)$,
{\it focus-focus} $(0,0,1)$.
\end{definition}

\begin{figure}[h]
\centering
\includegraphics{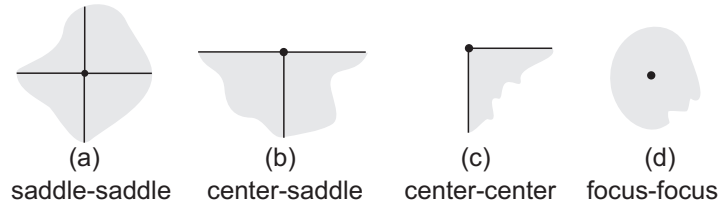}
\caption{Canonical bifurcation diagrams in the neighborhood of $\F(P)$ 
corresponding to functions $G_j$,
$n=2$.
The image of the momentum map is shaded gray.}
\label{FigBifDiagrCan}
\end{figure}

\begin{figure}[th]
\centering
\includegraphics{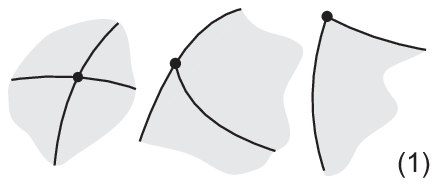}
\hspace*{1cm}
\includegraphics{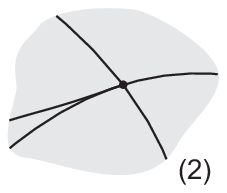}
\hspace*{1cm}
\includegraphics{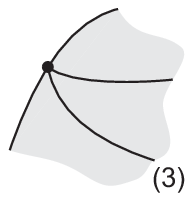}
\caption{Images $\F(K\cup \{P\})$ 
diagrams satisfying condition~(b) 
of Theorem~\ref{T1}, $n=2$.
The diagram (2) appears in the non-analytic case and (3) when the zero-rank point is
degenerate. The image of the momentum map is shaded gray.}
\label{FigBifDiagrGen}
\end{figure}

If $P$ is a non-degenerate
zero-rank singular point of an analytic IHS
then the bifurcation diagram around
$\F(P)$ {\it looks in the canonical way},
i.e.\ is
locally (at point $\F(P)$)
diffeomorphic to
the {\it canonical} bifurcation diagram
corresponding to functions $G_j$ \cite[1.8.4]{BF04}, \cite[p.9]{BO06}.
Figure~\ref{FigBifDiagrCan} 
shows these canonical bifurcation diagrams for $n=2$.
The canonical bifurcation diagram for Williamson type $(s,n-s,0)$
consists of $n$ hypersurfaces: $n-s$ hyperplanes
and $s$ half-hyperplanes.
For example, bifurcation diagrams on figure~\ref{FigBifDiagrGen}(1)
look in the canonical way.

Analogous statement exists if
we replace the bifurcation diagram by
the image $\F(K\cup\{P\})$
where 
$K$ is the set of all 
singularities of rank $1$.
The `canonical' image $\F(K\cup\{P\})$ 
for Williamson type $(s,n-s,0)$
consists of $n-s$ lines
and $s$ rays.

The converse is false: a point $P\in M$ 
can be a degenerate zero-rank
singular point such that the bifurcation diagram still looks in the canonical way
around $\F(P)$. A trivial example is as follows.
Denote $M:=\R^4$ with coordinates $(p_1,p_2,q_1,q_2)$,
$\omega:=dp_1\wedge dq_1+dp_2\wedge dq_2$,
$h_i:=p_i^4+q_i^4$ for $i=1,2$.
Then $(\R^4,\omega,h_1,h_2)$ is an IHS,
$P:=0\in \R^4$ is a degenerate zero-rank point,
but the bifurcation diagram consists of two lines
$x=0$ and $y=0$ on the plane $\R^2(x,y)$,
thus looks in the canonical way.

In this example we get $\D^2 h_i(P)=0$. 
A natural question arises: 
does the condition that the bifurcation diagram 
looks in the canonical way
{\it plus} some condition on $\D^2 h_i(P)$
(which holds for non-degenerate singularities
and which can be readily checked in real examples)
guarantee non-degeneracy of $P$?
Theorem~\ref{T1} gives the positive answer.

To prove that a singular point $P\in M$ 
is non-degenerate by definition,
one usually 
applies Lemma~\ref{LemCSA} below.
This
requires comparison of eigenvalues
which is a tricky computational task
(papers following this strategy are e.g. \cite{Mor02, BCRT}).
Theorem~\ref{T1} is intended to simplify
computation.
It is more effective
for IHSs of 2 and 3 degrees of freedom:
the geometric condition~(b)
can be effectively visualized then.


\begin{Th}
\label{T1}
Consider a completely integrable Hamiltonian system
$(M,\omega,h_1,\ldots,h_n)$.
Let
$\F:M\to\R^n$ be the momentum map
and $P\in M$
be a zero-rank singular point of the system.
Denote by $K$
the set of all singular points
of rank~1 in
a neighborhood of $P$.

If the following conditions hold, then $P$
is non-degenerate:
\\
(a) 
$\bigcap_{i=1}^n \Ker\D^2 h_i(P)=\{0\}$.
\\
(b) The image $\F(K\cup\{P\})$
contains $n$ smooth curves $\gamma_1,\ldots,\gamma_n$,
each curve having $P$ as its end point or its inner point.
\footnote{Figure~\ref{FigBifDiagrGen}~(1),(2),(3) shows examples for $n=2$.
}
The vectors
tangent to
$\gamma_1,\ldots,\gamma_n$
at $\F(P)$
are
independent in $\R^n$.
\\
(c) 
$K$ is a smooth submanifold of $M$ or, at least, $K\cup\{P\}$ coincides with the closure of
the set of all points $x\in K$
having a neighborhood
$V(x)\subset M$ for which $K\cap V(x)$
is a smooth submanifold of $M$.
\end{Th}

\begin{remark}
\label{remcond3}
Condition~(c) is very weak.
For example, it automatically holds
if the integrals $h_i$ are polynomials
(in a suitable system of local coordinates at point $P$)
because
in this case each $D_i$ is given by a system of
algebraic equations.
It also holds if $K$ consists of non-degenerate
singular points of rank~1
(in this case $K$ is smooth \cite[Proposition~1.18]{BF04}).
\end{remark}

\begin{remark}
\label{RemCond1}
By Lemma~\ref{LemMatr} below,
condition~(a)
is equivalent to the following
condition (a'):
{\it There exists a non-degenerate linear combination of forms
$\{\D^2h_i(P)\}_{i=1}^n$.}
\end{remark}

\begin{remark}[on the converse of Theorem~\ref{T1}]
\label{RemConverse}
In this remark we consider analytic IHSs for simplicity.
If $P$ satisfies Theorem~\ref{T1}, then it automatically
has elliptic-hyperbolic type,
i.e.\ its Williamson type is $(s,n-s,0)$ for some $s$,
see Definition~\ref{defwilliamsontype}.
Indeed, for a non-degenerate point of type $(s,2n-2k-s,2k)$,
$k>0$, the image $\F(K\cup\{P\})$ does not satisfy condition~(b)
by Theorem~\ref{ThNF}, see discussion above
and fig.~\ref{FigBifDiagrCan}(3).
So Theorem~\ref{T1} does not cover
focus-focus singularities.
The converse of Theorem~\ref{T1} is true for elliptic-hyperbolic singularities:
{\it Let $P$ be a non-degenerate zero-rank singular point of an
IHS, and suppose $P$
has Williamson type $(s,n-s,0)$ for some $s$.
Then it satisfies conditions (a),(b),(c)
of Theorem~\ref{T1}.}

This is well known.
Conditions (a)--(c) can be verified
in normal coordinates of Theorem~\ref{ThNF}.
Condition~(a) follows from the fact
that $\D^2 h_i(P)$ are independent.
The sets of critical points of rank $r$
for functions $h_i$ and $G_i$ coincide.
Hence condition~(c) follows \cite[Theorem~3]{BO06}; condition~(b)
also follows
as already stated above.
\end{remark}

\begin{remark}
In Condition~(c) of Theorem~\ref{T1} we do not demand that the image $\F(K\cup\{P\})$
coincides with the union of $\gamma_1,\ldots,\gamma_n$.
It may contain additional curves as on fig.~\ref{FigBifDiagrGen}(2),(3).
As discussed above, only $n$ curves appear in
the non-degenerate analytic case.
So Theorem~\ref{T1} implies the following interesting corollary. 
{\it If $P$ is a zero-rank singular point of an algebraic IHS
$(M,\omega,h_1,\ldots,h_n)$
and $\F(K\cup\{P\})$ contains more than $n$ curves with pairwise independent 
tangent vectors as on fig.~\ref{FigBifDiagrGen}(3)
then 
all linear combinations
of forms $\D^2h_1(P)$, $\D^2h_2(P)$
are degenerate.}
This can be observed in a wide range of
examples, for instance, in the Jukowsky integrable case
of rigid body dynamics \cite{Osh91, BF04}.
Here the assumption that IHS is algebraic is used
to guarantee
condition~(c), see Remark~\ref{remcond3}.
\end{remark}

\begin{remark}
\label{remsmooth}
In the $C^\infty$ case, Theorem~\ref{ThNF} is proved
for singularities of Williamson type $(s,n-s,0)$
\cite{Eli90, Mir03}
and very recently for focus-focus singularities $(0,0,1)$
\cite{SW11}. Remark~\ref{RemConverse} is true in the non-analytic case,
but now the bifurcation
diagram near the image $\F(P)$ of a non-degenerate singularity 
may split as shown on fig.~\ref{FigBifDiagrGen}(2)
(one curve splits into two curves 
with infinite order of tangency).
This example is found in \cite[1.8.4]{BF04}.
\end{remark}

We now turn to a criterion for non-degeneracy of $r$-rank
singularities. The definition
of non-degeneracy \cite[Definition~1.23]{BF04} is as follows.
\footnote{
This definition is equivalent to $P$ being a non-degenerate 
{\it zero-rank} singular point
of Marsden-Weinstein symplectic reduction
of the given system by 
the local action of $\R^{n-r}$ generated by flows of
Hamiltonian vector fields of $n-r$ independent integrals.
This helps to deduce Theorem~\ref{T2} easily from Theorem~\ref{T1}.
}

\begin{definition}
Let $(M,\omega,f_1,\ldots,f_n)$
be an IHS
and $P\in M$ be a singular point of rank $r$.
Find any regular linear change of integrals $f_1,\ldots,f_n$
so that the new functions, which we denote $h_1,\ldots,h_n$, satisfy the property:
$\D h_{r+1}(P)=\ldots=\D h_n(P)=0$. Consider the space
$L\subset T_PM$ generated by $\s h_1,\ldots, \s h_r$ and
its $\omega$-orthogonal complement $L'\supset L$.
Denote by $A_{r+1},\ldots,A_{n}$ the linear parts of vector fields
$\s h_{r+1},\ldots, \s h_n$. They are commuting operators in
$\sp(2n,\R)$. By \cite[Lemma~1.8]{BF04} the subspace $L$
belongs to the kernel of every operator $A_{r+1},\ldots,A_{n}$
and their image lies in $L'$. Thus they can be regarded as
operators on $L'/L$. By \cite[Lemma~1.9]{BF04} $L'/L$ admits a natural symplectic structure
and $A_{r+1},\ldots,A_{n}\in\sp(L'/L,\R)\cong \sp(2n-2r,\R)$.
The point $P\in M$ is called {\it non-degenerate} if
$A_{r+1},\ldots,A_{n}$ generate a Cartan subalgebra in $\sp(2n-2r,\R)$.
\end{definition}

\begin{remark}
Clearly, the definition does not depend on a regular 
$C^\infty(M)$-linear
change of the integrals. In Theorem~\ref{T2} we will consider integrals such that
that $\D h_{r+1}(P)=\ldots=\D h_n(P)=0$. To apply Theorem~\ref{T2} for a 
general integrable system
$(M,\omega,f_1,\ldots,f_n)$
it is sufficient to obtain integrals $h_i$ satisfying this property 
by a regular $C^\infty(M)$-linear change of $f_i$.
\end{remark}

\begin{Th}
\label{T2}
Consider a completely integrable Hamiltonian system
$(M,\omega,h_1,\ldots,h_n)$.
Let
$\F:M\to\R^n$ be the momentum map
and $P\in M$
be a singular point of rank $r$.
Denote by $K$
the set of all singular points
of rank~$r+1$ in
a neighborhood of $P$.
Suppose that $\D h_{r+1}(P)=\ldots=\D h_n(P)=0$
and $h_i(P)=0$ for all $i$.

If the following conditions hold, then $P$
is non-degenerate:
\\
(a) There exist 
a number $k\in\{r+1,\ldots,n\}$
and
a $(2n-2r)$-dimensional subspace $F\subset T_PM$
such that

($\mathit{a_1}$) $F\subset \bigcap_{j=1}^r\Ker\D h_j(P)$,

($\mathit{a_2}$) $F\cap \Lin\{\s h_1(P),\ldots,\s h_r(P)\}=\{0\}$ and

($\mathit{a_3}$)
$\bigcap_{i=r+1}^n\Ker\D^2 h_i(P)|_F=\{0\}$.
\\
(b) The intersection of the closure of $\F(K)\subset \R^n$ with
the submanifold $\{h_1=\ldots=h_r=0\}$
contains $n-r$ smooth curves,
each curve having $P$ as its end point or its inner point.
The vectors
tangent to
these curves
at $\F(P)$
are
independent in $\R^n$.
\\
(c) $K$ is an analytic submanifold of $M$ or, at least,
the closure of $K':=K\cap\{x\in M: h_1(x)=\ldots=h_r(x)=0\}$
coincides with the closure of
the set of all points $x\in K'$
having a neighborhood
$V(x)\subset M$ for which $K'\cap V(x)$
is a smooth submanifold of $M$.
\end{Th}

As in the case of Theorem~\ref{T1}, the converse of Theorem~\ref{T2} is true
for non-degenerate points of Williamson type $(s,n-r-s,0)$.


\section{Proofs of Theorems \ref{T1} and \ref{T2}}
\label{SecProofCrit}
We will need the following well-known lemmas.
We prove Lemma~\ref{LemMatr}
at the end of this section
since we do not have a reference for it.

\begin{Lemma}
\label{LemCSA}
{\rm (Cf. \cite[1.10.2]{BF04})}
A commutative subalgebra $K\subset sp(2n,\R)$
is a Cartan subalgebra
if and only if $K$ is $n$-dimensional and
it contains an element
whose eigenvalues are all different.
\end{Lemma}

\begin{Lemma}
{\rm (Cf. \cite[Lemma~2.20]{MDS98})}
Suppose
$A\in \sp(2n,\R)$ or $\sp(2n,\C)$.
If $\lambda\in\C$ is an eigenvalue of $A$,
then $-\lambda$ is also an eigenvalue of $A$.
\label{LemEig}
\end{Lemma}

\begin{Lemma}
\label{LemMatr}
Suppose
$A_1,\ldots,A_n\in GL(k,\R)$ 
commute pairwise. Then there exist
$\mu_i\in \R$
such that
$\Ker \sum_{i=1}^n \mu_i A_i=\bigcap_{i=1}^n\Ker A_i$.
\footnote
{
The following matrices:
$A_1=\left(
\begin{smallmatrix}
1	&0	\\
0	&0	\\
\end{smallmatrix}
\right)$,
$A_2=\left(
\begin{smallmatrix}
0	&1 \\
0	&0 \\
\end{smallmatrix}
\right)$
show that the 
commutativity condition is indeed necessary.}
\end{Lemma}

{\it Proof of Theorem \ref{T1}.}
{\it Step 1. Introducing new integrals.}
Denote $D_i:=\F^{-1}(\gamma_i)\cap K$.
Condition (b)
enables us to construct a new
set
$\{\h_i\}_{i=1}^n$
of independent commuting integrals
such that
$\h_j|_{D_i}\equiv 0$ for all
$i,j\in\{1,\ldots,n\}$, $i\neq j$.
Indeed, let
$g:\R^n\to\R^n$
be a diffeomorphism
taking $\gamma_i$ to the $i$-axis
and $\F(P)$ to $0\in \R^n$;
then define $\h_i:=gh_i$.
Below we work with the new integrals
$\h_i$.
Although the corresponding momentum maps
for $\{h_i\}$ and $\{f_i\}$ are different,
the critical set $K$
remains the same. Moreover, $\{\D^2\h_i(P)\}$
are obtained from $\{\D^2 h_i(P)\}$
by a regular linear change given by the operator
$\D g({\F(P)})$, so we can verify
Definition~1.1 for $\{\h_i\}$
as well as for $\{h_i\}$.
Below we write $\D^2\h_i$ instead of $\D^2\h_i(P)$
(and the same for other functions).
Denote
$$T_i:=
\bigcap_{\substack{j=1,\ldots, n\\ j\neq i}}
\Ker\D^2\h_j.$$
Denote by
$A_i\in \sp(T_PM)\cong \sp(2n,\R)$
the linear part of the vector field
$\s \h_i$
(equivalently, $A_i=\omega^{-1}\D^2\h_i$).
Clearly, $\Ker A_i=\Ker \D^2\h_i$
and $\{A_i\}_{i=1}^n$ commute pairwise.
Thus $T_j$ is $A_i$-invariant for each $i,j$.

{\it Step 2. Proof that $T_i\neq\{0\}$ for each $i$}.
\footnote{If we were given that $\overline{D_i}=D_i\cup\{P\}$
is a smooth submanifold,
then $T_i\neq \{0\}$ follows from the obvious inclusion
$T_P \overline{D_i}\subset T_i$.
We use condition~(c) in this step only.
}
Suppose to the contrary that $T_j=\{0\}$ for some $j\in\{1,\ldots,n\}$.
Then by Lemma~\ref{LemMatr}
some linear combination of 
$\{A_i\}_{i\neq j}$ is non-degenerate,
and thus the same combination of the 
forms $\{\D^2\h_j\}_{i\neq j}$ is non-degenerate.
Let $F$ be the linear combination
of functions $\{\h_i\}_{i\neq j}$ with the same coefficients.
We obtain:
($1^\circ$) $\D^2F$ is non-degenerate and
($2^\circ$) $F|_{D_j}\equiv 0$ since $\h_i|_{D_j}\equiv 0$ for $i\neq j$.
By ($1^\circ$) and the Morse lemma
there exists a punctured neighborhood $U'(P)\subset M$
of point $P$
such that
$\D F(x)\neq 0$
for all $x\in U'(P)$.
Now suppose $x\in D_j$ has a neighborhood
$V(x)$ such that $V(x)\cap D_j$
is a smooth submanifold.
By ($2^\circ$) we get
$\D(F|_{D_j})(x)=0$ for all $x\in U'(P)$,
meaning that $\D F(x)\perp T_xD_j$.
But $x$ is a point of rank 1,
so $\D F(x)$ and $\D \h_j(x)$ are linearly dependent.
Since $\D F(x)\neq 0$, this implies that
$\D \h_j(x)\perp T_xD_j$,
thus $\h_j|_{D_j}(x)=0$.
By (c), this holds for almost all $x\in U'(P)$
so $\h_j|_{D_j}\equiv\mathrm{const}$.
On the other hand,
$\h_j|_{D_j}$ is not a constant function
since the image $\h_j(D_j)$ is
a line segment and not a point.
This contradiction shows that $T_j\neq\{0\}$.

{\it Step 3. Proof that $\dim T_i\ge 2$ for each $i$.}
Suppose to the contrary that $\dim T_j=1$ for some $j$.
Without loss of generality, assume $j=1$.
Take $x\in T_1$, $x\neq 0$.
By definition, $A_i(x)=0$ for $i=2,\ldots, n$.
Then $A_1(x)\neq 0$, because
otherwise
$x\in  \bigcap_{i=1}^n\Ker A_i=\bigcap_{i=1}^n\Ker \D^2\h_i=\bigcap_{i=1}^n\Ker \D^2h_i$,
which contradicts to condition~(a).
But $T_1$ is $A_1$-invariant, and
we obtain $A_1(x)=\lambda x$ for some $\lambda\neq 0$.
Lemma~\ref{LemEig} implies that $(-\lambda)$ is also an eigenvalue of $A_1$,
meaning that there exists $y\in T_PM$, $y\neq 0$,
such that $A_1(y)=-\lambda y$. 
The subspace $L:=\Lin (\{x,y\})$ is symplectic
and $A_i$-invariant
for each $i=1,\ldots,n$. In its basis
$\{x,y\}$ we get
$$
A_1|_L=\lambda
\begin{pmatrix}
1 & 0\\
0 & -1
\end{pmatrix}
\quad
\mbox{and}
\quad
A_i|_L=
\begin{pmatrix}
0 & b_i\\
0 & c_i
\end{pmatrix}
\quad
\mbox{for}\ 
i\ge 2.
$$
Since $A_1$ commutes with $A_i$, we obtain
that $b_i=0$ for $i\ge 2$.
Since $A_i|_L\in sp(L)$, we obtain 
that $c_i=0$ for $i\ge 2$. Consequently, $A_i(y)=0$, $i\ge 2$.
By definition this means $y\in T_1$.

{\it Step 4. Proof that $\dim T_i=2$
and $\bigoplus_{i=1}^n T_i=T_PM$.}
By condition~(a),
any $n$ non-zero vectors $v_i\in T_i$, $1\le i\le n$,
are independent.
(Indeed, suppose to the contrary that $v_1$
is a linear combination of $\{v_2,\ldots,v_n\}$.
Then by construction
$v_1\in \bigcap_{i=1}^n\Ker \D^2\h_i=\bigcap_{i=1}^n\Ker \D^2h_i$,
which contradicts to condition~(a).)
Combining this with Step~3
we obtain that
$\dim T_i=2$
and
$\bigoplus_{i=1}^n T_i=T_PM$
for each $i\in\{1,\ldots,n\}$.

{\it Step 5. Final step.}
By construction,
$\Ker A_i=\Ker \D^2\h_i=
\bigcup_{j\in\{1\ldots n\}\setminus\{i\}}
{T_j}$.
This means that
for all $i,j\in\{1,\ldots,n\}$, $i\neq j$,
we obtain 
$A_i|_{T_j}\equiv 0$.
Condition~(a) now implies that
$\Ker A_i|_{T_i}=\{0\}$.
So
the eigenvalues of $A_i|_{T_i}$
are $\{\pm \lambda_i\neq 0\}$
for some $\lambda_i\in\C$.
Let us prove that $P$ is non-degenerate.
Clearly, $\{A_i\}_{i=1}^n$ are independent.
The eigenvalues of
a linear combination
$\sum_{i=1}^n \mu_i A_i$
are $\{\pm \mu_i\lambda_i\}_{i=1}^n$
which are obviously all different for well-chosen coefficients $\mu_i$.
Thus $P$ is non-degenerate by Definition~1.1,
Lemma~\ref{LemCSA} and the argument in Step~1. 
Proof of Theorem~\ref{T1} is 
finished.~$\blacksquare$

\smallskip
{\it Proof of Theorem~\ref{T2}.}
By the Darboux theorem, we can complete functions
$p_1:=h_1,\ldots,p_r:=h_r$ up to a coordinate system
$\{p_i,q_i\}_{i=1}^n$ at point $P$ such that
$\{p_i,p_j\}=0$, $\{p_i,q_j\}=\delta_{ij}$ for all
$1\le i,j\le n$.
Denote $\Pi:=\Lin\{\dd p_i,\dd q_i\}_{i=r+1}^n\subset T_PM$.
Consider the symplectic submanifold $Q\subset M$ in a neighborhood of $P$
given by equations $\{p_i=0,q_i=0\}_{i=1}^r$; then $T_PQ=\Pi$.
By Definition~1.2, $P$ is non-degenerate
if the restricted operators $\{\omega^{-1}\D^2h_{r+1}|_\Pi,\ldots,\omega^{-1}\D^2h_n|_\Pi\}$
generate a Cartan subalgebra of $\sp(2n-2r,\R)$.
Clearly, this is equivalent to $P$ being a non-degenerate
zero-rank singular point of the reduced IHS $(Q,\omega|_Q,\{h_i|_Q\}_{i=r+1}^n)$.
We can apply Theorem~\ref{T1} to this reduced system by
verifying the three conditions of Theorem~\ref{T1}.

By~(a$_3$) there is a linear combination $H$
of $h_{r+1},\ldots,h_n$ such that $\D^2 H|_F$ is non-degenerate.
By (a$_1$) $F\subset\Lin (\Pi\cup\{\dd q_i\}_{i=1}^r)$;
by (a$_2$) the projection $F\stackrel{\mathrm{pr}}\to\Pi$ has zero kernel
and is an isomorphism since $\dim\ F=\dim\ \Pi$.
Since $h_i$ commute, it follows that $\{h_i\}_{i=r+1}^n$ do not depend on
$\{q_1,\ldots,q_r\}$, so
$\D^2H(v)=\D^2h_k(\mathrm{pr}\,v)$ for $v\in \Lin (\Pi\cup\{\dd q_i\}_{i=1}^r)$.
Together with (a$_3$) this implies that
$\D^2H|_\Pi$ is non-degenerate.
Condition (a) of Theorem~\ref{T1} is verified.

Let $\tilde K$ and $\tilde\F$ 
denote respectively the set of 1-rank points near $P$ and the momentum map of the
restricted system.
Condition~(b) of Theorem~\ref{T1} follows from the given condition~(b)
because $\tilde\F(\tilde K)=\F(K)\cap\{x\in M:\ h_1=\ldots=h_r=0\}$.
Condition~(c) of Theorem~\ref{T1} follows from the given condition~(c).
Indeed, $\tilde K=K\cap Q$ and
since 
all gradients $\{\D h_i\}_{i=1}^n$ are independent of $\{q_i\}_{i=1}^r$,
$K'$ is a cylinder over $\tilde K$.
So if $K$ is an analytic submanifold or $K'$ is `almost everywhere regular' in the sense
of condition (c) then $\tilde K$ also `almost everywhere regular', i.e.\ satisfies
condition (c) in 
Theorem~\ref{T1}.~$\blacksquare$

\smallskip
{\it Proof of Lemma~\ref{LemMatr}.}
Let us first prove the lemma for
$n=2$; denote $A:=A_1$, $B:=A_2$.
Consider a basis $(e_1,\ldots,e_k)$
for $\R^k$ such that $(e_1,\ldots,e_j)$
spans $\Ker A$ for some $j$.
In this basis we get
$A=
\left( \begin{smallmatrix}
0_{j{ \times } j} & A''\\
0 & A'
\end{smallmatrix} \right)$
and
$B=
\left( \begin{smallmatrix}
B'''& B''\\
0 & B'
\end{smallmatrix} \right)$.
Here $0_{j{ \times } j}$ and $B'''$
are $j{\times} j$-matrices
and $A',B'$ are $(k-j){\times}(k-j)$-matrices.
By construction $A'$ is non-degenerate.
Clearly $\Ker(A+\varepsilon B)=\Ker A\cap\Ker B$ 
for sufficiently small $\varepsilon$.

The general case is proved by induction on $n$.
Let us prove the step. Given $A_1,\ldots,A_n$,
we can find by the induction hypothesis a linear combination $B$
of $A_1,\ldots A_{n-1}$ whose kernel is
$\bigcap_{i=1}^{n-1}\Ker A_i$.
By the $n=2$ case,
there is a linear combination of $B$ and $A_n$
whose kernel is $\Ker B\cap\Ker A_n=\bigcap_{i=1}^{n}\Ker A_i$.~$\blacksquare$

\section{Application to the classical and quantum Manakov top}
\label{SecManakov}
\subsection{A short introduction to the Manakov top system}
\label{SubSecDefManakov}
The Manakov top integrable
system
(also known as the geodesic flow on $so(4)$
and the 4-dimensional rigid body)
was introduced in \cite{Man76}.
Oshemkov \cite{Osh87, Osh91}
\footnote{The result from these references are
also found in \cite[vol.~2, 5.10]{BF04}}
studied the topology and bifurcation diagrams of
the system; we reproduce the bifurcation diagrams
below.
For certain parameters,
the Manakov top
contains a focus-focus point.
The corresponding {\it Hamiltonian  monodromy} \cite{Dui80}
was calculated by Audin \cite{Au02}
using algebraic technique
which allowed not to check non-degeneracy of the point.


Let us recall the Manakov top system following \cite{Osh91}.
Consider $\R^6$ with coordinates $p_1,p_2,p_3$,
$m_1,m_2,m_3$.
Define the Lie-Poisson bracket on $\R^6$:
$$
\{m_i,m_j\}=\epsilon_{ijk}m_k,\quad
\{m_i,p_j\}=\epsilon_{ijk}p_k,\quad
\{p_i,p_j\}=\epsilon_{ijk}m_k.
$$
Here $\epsilon_{ijk}=(i-j)(j-k)(k-i)$.
This bracket
has two Casimir functions
$$f_1=m_1^2+m_2^2+m_3^2+p_1^2+p_2^2+p_3^2,\quad
f_2=m_1p_1+m_2p_2+m_3p_3.$$
Fix three numbers $0<b_1<b_2<b_3$.
Functions
\begin{gather*}
h_1=b_1m_1^2+b_2m_2^2+b_3m_3^2-(b_1p_1^2+b_2p_2^2+b_3p_3^2),\\
h_2=(b_1+b_2)(b_1+b_3)p_1^2+(b_2+b_1)(b_2+b_3)p_2^2+(b_3+b_1)(b_3+b_3)p_3^2
\end{gather*}
commute with respect to the defined bracket
and thus define an IHS on a symplectic leaf
$$M^4_{d_1,d_2}:=\{x\in\R^6: f_1(x)=d_1,\ f_2(x)=d_2\}$$
of the Lie-Poisson bracket, $|2d_2|<d_1$.
This system is called the Manakov top.
Its parameters are $(b_1,b_2,b_3,d_1,d_2)$.

\begin{figure}[h]
\centering
\includegraphics{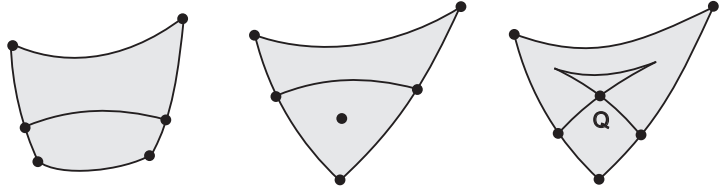}
\caption{Three types of generic bifurcation diagrams of the Manakov top. 
Point $Q$ is the image of two saddle-saddle singularities.}
\label{FigBifDiagrManakov}
\end{figure}

For a certain (open) set of parameters $b_i,d_i$,
the bifurcation diagram has one of the three types
shown on fig.~\ref{FigBifDiagrManakov};
see \cite{Osh91} for details.
The diagram of the third type
separates the image of the momentum map
into three domains.
The $\F$-preimage of each 
point of the inner domain
consists of 4 tori.
The preimage of each 
point of the two other domains
consists of 2 tori.
Let $Q$ be the intersection point of the
two inner curves on the bifurcation diagram,
see fig.~\ref{FigBifDiagrManakov}.
The preimage $\F^{-1}(Q)$ contains two zero-rank points \cite{Osh91}.
It is natural to expect that they
are non-degenerate saddle-saddle singularities.
The proof becomes simple with the help of Theorem~\ref{T1}.

\subsection{Non-degenerate singularities of the Manakov top}
The explicit parameters of the Manakov top under which
the system contains degenerate singularities
were recently obtained in \cite[Theorem~5.3]{BCRT}, 
cf.\ \cite{BC}.
In the following proposition, the
description of
non-degenerate singularities is
very natural: it essentially says
that all degenerate singularities are easily seen
to be degenerate by looking at the bifurcation diagrams.
As already mentioned, the proof of
Proposition~\ref{ThNondegen} using Theorem~\ref{T1}
is considerably shorter than the proofs in \cite{BCRT}.
\footnote{
In the proof of Proposition~\ref{ThNondegen} we essentially determine
the parameters $(b_i,d_i)$
which contain degenerate singularities.
They seem to agree with those from~\cite{BCRT},
although that paper uses different notation.
Also, Proposition~\ref{ThNondegen}
could be deduced from \cite{BCRT}
and even is in part implicitly stated there,
see~\cite[text after Theorem~5.3]{BCRT}.
}
Recall the Williamson type of a non-degenerate
singularity was introduced in Definiton~\ref{defwilliamsontype}.

\begin{Proposition}
\label{ThNondegen}
Let $P\in M$ be a zero-rank singular point of the Manakov top
with parameters $(b_i,d_i)$.
Then $P$ is non-degenerate and not of focus-focus type 
if and only if 
for each set
of parameters $(b_i',d_i')$
sufficiently close to $(b_i,d_i)$,
the bifurcation diagram of the Manakov top
with parameters $(b_i',d_i')$
can be transformed by a diffeomorphism
of a neighborhood of $\F(P)$ 
to one of the three diagrams shown on
fig.~\ref{FigBifDiagrCan}(a,b,c).
\footnote{As previously mentioned, Theorem~\ref{T1}
does not cover focus-focus singularities, so we have to exclude
them from this proposition as well.}
\end{Proposition}

(The `only if' part of Proposition~\ref{ThNondegen} is trivial.)
Degenerate singularities thus do not appear 
when the bifurcation diagram 
has one of the generic types
shown on fig.~\ref{FigBifDiagrManakov},

\begin{Corollary}
\label{ThNondegenSaddle}
Let $P\in M$ be a zero-rank singular point of the Manakov top
with parameters $(b_i,d_i)$.
Then $P$ is a non-degenerate saddle-saddle singular point
if and only if 
the bifurcation diagram of the Manakov top
with parameters $(b_i,d_i)$
can be transformed by a diffeomorphism
of a neighborhood of $\F(P)$ 
to the diagram shown on
fig.~\ref{FigBifDiagrCan}(a).

In this case, $\F^{-1}(\F(P))$
contains two zero-rank points, both
of saddle-saddle type.
\end{Corollary}

{\it 
Proof of Corollary~\ref{ThNondegenSaddle}
modulo \prop~\ref{ThNondegen}.}
By looking at
the types of bifurcation diagrams in \cite{Osh91}
it easily seen that hypothesis of the 
Corollary~\ref{ThNondegenSaddle}
is stable under parameter perturbation
and thus implies the hypothesis
of
\prop~\ref{ThNondegen}.
The fact that $\F^{-1}(\F(P))$ contains two zero-rank points
is proved in \cite{Osh91} and is easy;
it also follows from the proof of \prop~\ref{ThNondegen}.~$\blacksquare$

\smallskip
For example, if $Q\in\R^2$
is the point from fig.~\ref{FigBifDiagrManakov} or fig.~\ref{figManakovBifDiagrSym},
the two zero-rank points in the preimage $\F^{-1}(Q)$
are nondegenerate and of
saddle-saddle type.

\begin{remark}
There are higher-dimensional versions of
the Manakov top system, called the $n$-dimensional
rigid body. For $n\ge 5$ it should be explored using a different
approach because it is hard 
to study the bifurcation diagrams of this system.
Remarkably, an approach using the bi-Hamiltonian structure
provides the complete answer (A.~Izosimov, preprint).
\end{remark}

\subsection{Semilocal structure of saddle-saddle singularities
of the Manakov top}
Recall that an IHS
$(M^4,\omega,f_1,f_2)$
defines the singular {\it Liouville foliation}
on $M$
whose fibers are 
common level sets of functions $(f_1,f_2)$,
i.e. level sets of the momentum map $\F$.
Regular fiber of this foliation is a disjoint union of tori 
(under certain assumptions which hold in the Manakov top)
\cite{Arn78}.

\begin{definition}
We will call a diffeomorphism preserving Liouville foliation a
{\it Liouville equivalence} or a ($\F$-)fiberwise diffeomorphism.
\end{definition}

In \prop~\ref{ThSemiprod} below we describe 
{\it semilocal structure} of the saddle-saddle
singularities of the Manakov top, 
i.e.\ describe the (singular) Liouville foliation
on $\F^{-1}(V)$ where $V\subset \R^2$
is a small neighborhood of $Q$.
\footnote{
The word `semilocal' is used since 
the preimage $\F^{-1}(V)$, even $\F^{-1}(Q)$,
is not at all local, i.e.\ 
does not belong to small neighborhood in $M^4$.
It contains two distant zero-rank singularities.}

\begin{figure}[h]
\centering
\includegraphics{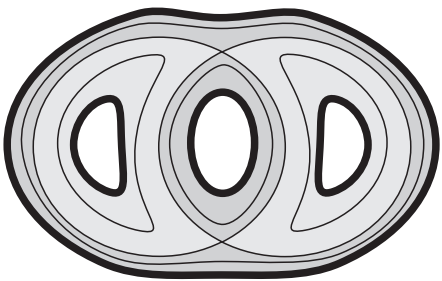}
\caption{The fibered 2-manifold $\FC$.}
\label{figC2flat}
\end{figure}

To state \prop~\ref{ThSemiprod}, we have to introduce some notation
(cf. \cite{BF04, Kudr99}).
Let
$\FC$
be the fibered 2-manifold with boundary
shown on fig.~\ref{figC2flat}.
Formally,
$\FC$
is the preimage $h^{-1}(-\varepsilon,\varepsilon)$
of a certain Morse function $h:\R^2\to\R$
having two singular points at one critical value $0$.
Level sets
of $h$ define
the singular fibration on $\FC$.
Two shades on fig.~\ref{figC2flat} show the areas below and over the critical
value of $h$.
A regular fiber 
on $\FC$
is a disjoint union of two circles.
The circles in $\bd \FC$
are distributed between two fibers.
The direct product
$\FC\times \FC$ is a 4-manifold with boundary
equipped with the product fibration.
\footnote{This fibration comes from an
integrable system on $\FC\times \FC$
\cite[9.6]{BF04} so can be called Liouville foliation.}
Its regular fiber is a disjoint union of 
four tori.
Let $\alpha$
be rotation by $180^\circ$,
the free fiberwise involution on $\FC$.
The involution $(\alpha,\alpha)$
preserves fibration on $\FC\times \FC$
and thus defines the fibered 4-manifold
$(\FC\times \FC)/(\alpha,\alpha)$.

\begin{Proposition}
\label{ThSemiprod}
Let $Q\in \R^2$ be the point on the bifurcation diagram of the Manakov top
as on fig.~\ref{FigBifDiagrManakov} or fig.~\ref{figManakovBifDiagrSym}
and $V$ its neighborhood such that $\F^{-1}(V)$
retracts onto $\F^{-1}(Q)$.
Then $\F^{-1}(V)$ is Liouville equivalent to 
$(\FC\times\FC)/(\alpha,\alpha)$.
\end{Proposition}

Fomenko and his collaborators
obtaied a complete list of the Liouville equivalence
classes of neighborhoods of 
singular Liouville fibers containing two non-degenerate saddle-saddle
singular points for all integrable systems
with two degrees of freedom:
\cite[9.6, Tables~9.1 and~9.3]{BF04},
compare \cite{Bol91, Mat93}.
Since $Q$ is non-degenerate by
Corollary~\ref{ThNondegenSaddle},
$\F^{-1}(V)$
is Liouville equivalent to one of the 39 items
from these tables.
To prove \prop~\ref{ThSemiprod},
we just have to identify the correct item.
It is very easy, see the proof in \S\ref{SecProofManakov}.

Note there is a general
theorem by Nguen~Tien~Zung
stating that
all neighborhoods of Liouville fibers containing saddle-saddle singularities
can be obtained as a quotient of 
a direct product of certain fibered 2-manifolds
\cite{NTZ95}.

\subsection{Action variables around saddle-saddle singularities
of the Manakov top and relation to the quantum Manakov top}
Our last goal is to describe the structure
of action variables around the singular fiber
containing saddle-saddle singularities of
the Manakov top. First, we recall \cite[Appendix A]{SiZh07} 
that under some parameters,
the Manakov top has some symmetries
and satisfies the following Condition~\ref{CondSym}.
Recall that $\F$ is the momentum map $M\to\R^2$,
where $(M,\omega)$ is the phase space of the Manakov top system.
Each regular fiber of $\F$ is a disjoint union of 2 or 4 tori.

\begin{condition}
\label{CondSym}
Every Liouville torus can be mapped onto any
other torus on the same regular $\F$-fiber via an $\F$-preserving 
symplectomorphism of $(M,\omega)$.
\end{condition}

\begin{figure}[h]
\centering
\includegraphics{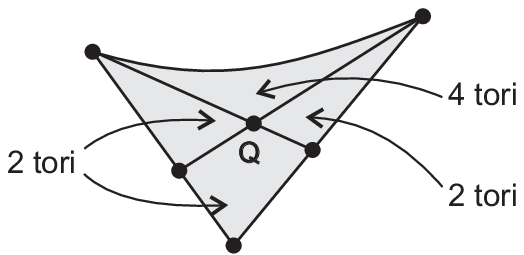}
\caption{Bifurcation diagram of the Manakov top system
satisfying Condition~\ref{CondSym}}
\label{figManakovBifDiagrSym}
\end{figure}

In notation of
Subsection~\ref{SubSecDefManakov},
Condition~\ref{CondSym} is satisfied if $d_2=0$.
The 
group of $\F$-preserving symplectomorphisms
which ensures Condition~\ref{CondSym} is generated by
$(m_i,p_i)\mapsto(-m_i,p_i)$
and
$(m_i,p_i)\mapsto(m_i,-p_i)$.
For $d_2=0$, the bifurcation diagram of the Manakov top looks as shown on
fig.~\ref{figManakovBifDiagrSym}.

Condition~\ref{CondSym} implies that action variables 
on a regular torus are the same on the other tori
of the same $\F$-fiber, which means they can be regarded as functions
over the image of the momentum map, a domain in $\R^2$.

In the following proposition part~(c) is most interesting
in the context 
of the quantum Manakov top.
It
describes up to homeomorphism
the  $\F$-image of 
`Bohr-Sommerfeld' tori of the Manakov top, i.e. those
tori on which the values of action
variables belong to $2\pi h\Z$, $h\in \R$.

This proposition is an easy corollary of
the purely topological 
\prop~\ref{ThSemiprod} and is proved in \S\ref{SecProofManakov}.

\begin{figure}[h]
\centering
\includegraphics{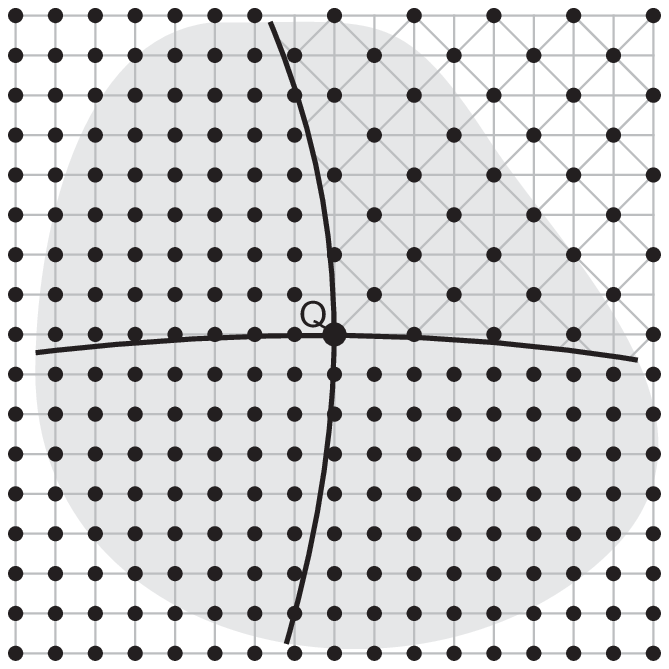}
\hspace{0.5cm}
\includegraphics{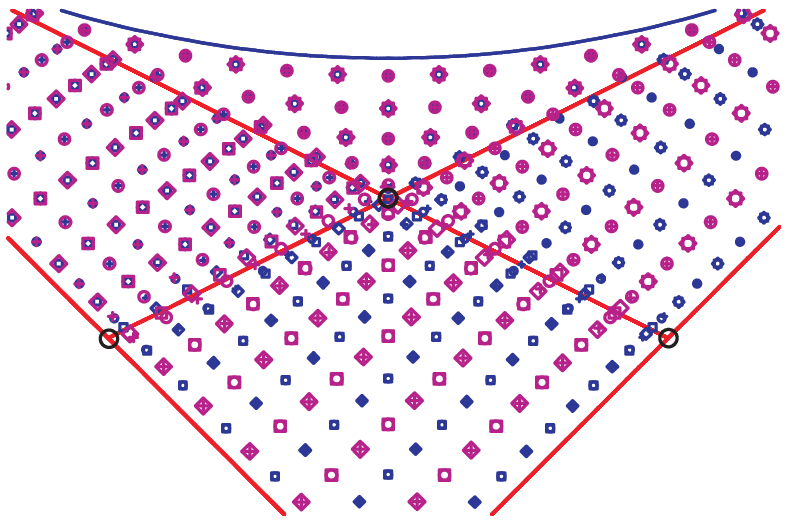}
\caption{Left: an example of the lattice $\psi\F(L_h)$ from \prop~\ref{ThLattice}.
The two curves are the $\psi$-image of the bifurcation diagram.
The shaded area is $\psi(V)$.
Right:
a  \textsc{reproduction of} \cite[figure  13]{SiZh07},
joint spectrum of the quantum Manakov top
computed by Sinitsyn and Zhilinskii.
}
\label{figManakovLattice}
\end{figure}

\begin{Proposition}
\label{ThLattice}
Consider the Manakov top system $(M,\omega,h_1,h_2)$ 
with parameter $d_2=0$ (i.e. satisfying Condition~\ref{CondSym})
and containing a saddle-saddle singularity $P$.
Let $V\subset \R^2$ be
a small neighborhood of $Q:=\F(P)$ 
such that $U:=\F^{-1}(V)$
retracts onto $\F^{-1}(Q)$. 
There is a 1-form
$\theta$ on $U$ such that $d\theta=\omega|_U$ 
and two continuous functions $a_1,a_2:V\to \R$ such that:

(a) 
$a_1,a_2$ are smooth at regular values of $\F$.
For each Liouville torus $T\subset U$, there is
a basis $(\rho_1,\rho_2)$ of $H_1(T;\Z)$
such that 
$$
a_1(\F(T))=\int_{\rho_1}\theta,
\quad
a_2(\F(T))=\int_{\rho_2}\theta
$$
if $\F^{-1}(\F(T))$ consists of two tori
and
$$
\frac{1}{2}(a_1-a_2)(\F(T))=\int_{\rho_1}\theta,
\quad
\frac{1}{2}(a_1+a_2)(\F(T))=\int_{\rho_2}\theta
$$
if $\F^{-1}(\F(T))$ consists of four tori.
\footnote{It means that the globally defined functions $a_1,a_2$
are action variables up to a linear change.}

(b) 
The map $\psi:=(a_1;a_2)$
is a homeomorphism from
$V$ to a neighborhood
of $(0;0)\in \R^2$
taking $Q$ to $(0;0)$.
Here $\R^2$ is equipped with standard coordinates $(x,y)$.
The $\psi$-image of the bifurcation diagram is
a union of two $C^1$-curves intersecting at $(0;0)$.
At this point, one of these curves is tangent
to the $x$-axis,
and the other one to the $y$-axis.
Also, 
$\psi$ is $C^\infty$ outside the bifurcation diagram.

(c) Let $L_h$, $h\in \R_+$, 
be the union of all Liouville tori in $U$ satisfying the following
condition:
the values of all action functions (with respect to the 1-form $\theta$)
on the torus belong to $2\pi h\Z$.
\footnote{
If $T$ is the torus in question,
this is equivalent by part (a) to
$a_1(\F(T))$, $a_2(\F(T))\in 2\pi h\Z$
if $\F^{-1}(\F(T))$ consists of 2 tori
and
$1/2(a_1-a_2)(\F(T))$, $1/2(a_1+a_2)(\F(T))\in 2\pi h\Z$
if $\F^{-1}(\F(T))$ consists of 4 tori.
}
For each $h\in\R_+$ the homeomorphism $\psi$ takes the set $\F(L_h)$
to the following set 
(which is a subset of the \emph{straight} lattice
$2\pi h\Z\times 2\pi h\Z$; see an example on fig.~\ref{figManakovLattice} left):
$$\{(x,y)\in \psi(V)
\quad
\mbox{such that}
\quad
\begin{cases}
x,y\in 2\pi h\Z,& \mbox{if } \F^{-1}(\phi^{-1}(x,y)) \mbox{ consists of 2 tori }
\\
x-y,x+y\in 4\pi h\Z, & \mbox{if } \F^{-1}(\phi^{-1}(x,y)) \mbox{ consists of 4 tori}
\end{cases}
\quad
\}.
\footnote{Recall that the bifurcation diagram splits $V$ into
four domains. On one of these domains, the $\F$-preimage
of a point consists of 4 tori, and on the other ones it 
consists of 2 tori.}
$$
\end{Proposition}

Part (c) is most interesting in the context of quantization
of the Manakov top system. 
Roughly speaking, it predicts the qualitative
view of the joint spectrum lattice
of a quantized Manakov top.

For the quantum Manakov top described in \cite{KK91},
the joint spectrum
of the two quantum operators was numerically computed
and visualized by Sinitsyn and Zhilinskii
\cite[figures 1 and 13]{SiZh07}.
For convenience, we reproduce 
\cite[figure 13]{SiZh07}
on fig.~\ref{figManakovLattice} right.
By Proposition~\ref{ThLattice},
the `Bohr-Sommerfeld lattice'
$\F(L_h)$
of the Manakov top
up to homeomorphism
looks as on
fig.~\ref{figManakovLattice} left.
The reader is invited to compare
figs.~\ref{figManakovLattice} left and right:
they are very similar!

Fig.~\ref{figManakovLattice} (left) grasps
the main features of the lattice
from \cite{SiZh07}. Note
that our figure is obtained by general arguments,
without any computation.
An analogue of \prop~\ref{ThLattice}
is true for other integrable systems
having saddle-saddle singularities of
the same type, including the Clebsch system.

Proposition~\ref{ThLattice}
describes the lattice $\F(L_h)$
`up to homeomorphism'.
There are general
results stating that $\F(L_h)$
(or its modification, e.g. a Maslov-type correction)
approximates the spectral lattice
of the quantum system for different quantization
schemes including
Toeplitz quantization \cite{Ch03},
Maslov asymptotic quantization \cite{KK93},
pseudo-differential quantization 
(the first two are applicable to the Manakov top).
Unfortunately, 
the author was
not able to find any general result of this kind
in the framework of quantization used in \cite{SiZh07}.
Here we {\it do not prove} that $\F(L_h)$
does indeed approximate the spectrum of the quantum Manakov top
from \cite{SiZh07}. 
Discussion above shows this is very likely to be true.

\begin{remark}
When we say that
fig.~\ref{figManakovLattice}
is similar to \cite[figure 13]{SiZh07},
we ignore
different symmetry types of the eigenvalues
pictured by different shapes and colors 
in \cite[figure  13]{SiZh07}
(i.e. consider all the points on these figures as black points).
The author is grateful to Professor B.I.~Zhilinskii
for indicating that
there is an important feature of rearrangement
of different types of eigenvalues
near the bifurcation diagram.
It would be very interesting to find
a classical description of this phenomenon
as well. 
\end{remark}

\section{Proofs of \prop s \ref{ThNondegen}, \ref{ThSemiprod} and \ref{ThLattice}}
\label{SecProofManakov}
{\it Proof of \prop~\ref{ThNondegen}.}
Denote $M:=M_{d_1,d_2}$.
We will check the three conditions of Theorem~\ref{T1}.
Condition~(b) is obvious.
Condition~(c) holds automatically, see Remark~\ref{remcond3}.
It is left to check condition~(a). We will check
the equivalent condition (a') from
Remark~\ref{RemCond1} instead.
Denote $H_i=h_i|_M$.
For each $i=1,2$ we get $\D H_i(P)=0$.
This is equivalent to
the fact that
$$\D h_i(P)=\lambda_i \D f_1(P)+\mu_i\D f_2(P)$$
for some $\lambda_i,\mu_i\in\R$.
It is easy to check \cite{Osh91}
that the equation $\D H_1(P)=0$
has exactly twelve solutions for $P\in M$:
\begin{gather*}
(\pm A,0,0,\pm B,0,0),\quad (\pm B,0,0,\pm A,0,0),\\
(0, \pm A,0,0,\pm B,0),\quad (0,\pm B,0,0,\pm A,0),\\
(0,0, \pm A,0,0,\pm B),\quad (0,0,\pm B,0,0,\pm A),
\end{gather*}
where $2A=\sqrt{d_1+2d_2}+\sqrt{d_1-2d_2}$,
$2B=\sqrt{d_1+2d_2}-\sqrt{d_1-2d_2}$.
At these points we also get $\D h_2(P)=0$,
so they are of zero rank.
We can assume that
$P=(\pm A,0,0,\pm B,0,0)$
(other points are considered analogously).
Let us find a combination $h_1+\alpha h_2$
such that $\D(h_1+\alpha h_2)(P)=\beta \D f_1(P)$.
Easy calculation shows that
$$
\begin{pmatrix}
\D h_1(P)\\
\D h_2(P)\\
\D f_1(P)
\end{pmatrix}
=
\begin{pmatrix}
2b_1m_1,&0,&0,&-2b_1p_1,&0,&0\\
0,&0,&0,&2(b_1+b_2)(b_1+b_3)p_1,&0,&0\\
m_1,&0,&0,&p_1,&0,&0\\
\end{pmatrix}
$$
so we can take $\beta=2b_1$, $\alpha=2b_1/(b_1+b_2)(b_1+b_3)$.
Let us prove that
$\D^2(H_1+\alpha H_2)(P)$ is
a non-degenerate
form on $T_PM$.
Clearly
$$\D^2(H_1+\alpha H_2)(P)=
(\D^2 (h_1+\alpha h_2-\beta f_1)(P))|_{T_PM}.$$
In the 
basis $(\bd/\bd p_2,\bd/\bd p_3,\bd/\bd q_2,\bd/\bd q_3)$
for $T_PM$
we get
$$
\D^2(H_1+\alpha H_2)(P)=
2\,\mathrm{diag}\,(b_2-b_1,b_3-b_1,c/(b_1+b_3),c/(b_1+b_2))
$$
where $c=b_1b_2+b_1b_3-b_2b_3-b_1^2$.
If $c\neq 0$, then condition~(a) is satisfied
and $P$ is non-degenerate by Theorem~\ref{T1}.

Suppose $c=0$; we will come to a contradiction.
Let $\{b_1',b_2',b_3'\}$
be some parameters
close to $\{b_1,b_2,b_3\}$.
Let $h_1',h_2'$ be the integrals of the system
corresponding to parameters $\{b_i',d_1,d_2\}$
and $H_i'=h_i'|_M$.
Define $\alpha '$, $c'$ analogously to $\alpha$, $c$
replacing $b_i$ by $b_i'$.
We can choose $b_i'$ such that $c'\neq 0$,
hence condition~(a) of Theorem~\ref{T1} is satisfied for the
system with parameters $b_i'$.
By the hypothesis,
the bifurcation diagram
for
$b_i'$ also satisfies condition~(b).
Thus $P$ is non-degenerate for the system with parameters $b_i'$.
Moreover, by the hypothesis and Theorem~\ref{ThNF}
point $P$ has the same Williamson type (see \S\ref{SecInt})
for each $b_i'$. Thus
any linear combination of
Hessians of the
integrals, including combination
$
\D^2(H_1'+\alpha 'H_2')(P)
$,
has the same signature
for each $b_i'$.
On the other hand,
$c'$ can be of arbitrary sign when $b_i'$
are arbitrarily close to $b_i$.
Thus there are two sets of parameters 
$b_i'$ arbitrarily close to $b_i$
such that
$
\D^2(H_1'+\alpha ' H_2')(P)
$
has different signatures.
This contradiction proves that $c\neq 0$.~$\blacksquare$


\smallskip
{\it Proof of \prop~\ref{ThSemiprod}.}
By Corollary~\ref{ThNondegenSaddle} and \cite[Theorems~9.7,~9.8]{BF04},
$\F^{-1}(V)$
is Liouville equivalent to one of the 39 items
from \cite[9.6, Table~9.1]{BF04}.
It is easy to identify the correct item.
We know that
the numbers of tori in the preimage of a point in $V$
are 2/2/2/4 depending on one of the four domains containing the point.
The only two items in the table \cite[Table~9.1]{BF04}
satisfying this condition have numbers 12 and 17.
However, item 12 is different because it contains
a non-orientable separatrix, and 
by \cite{Osh91} the Manakov top does not.
Thus $\F^{-1}(V)$ is Liouville equivalent
to item 17 from \cite[Table~9.1]{BF04}
which corresponds by \cite[Table~9.3]{BF04}
to $(\FC\times \FC)/(\alpha,\alpha)$.~$\blacksquare$

\smallskip
{\it Proof of Proposition~\ref{ThLattice}.} 
{\it Step 1. Lift to the direct product.}
The symplectic form $\omega$
is exact on $U$ because
$U$ retracts onto the fiber
$\F^{-1}(\F(P))$
which is Lagrangian.
Let $\theta$ be a 1-form on $U$
such that $d\theta=\omega$.
Recall Proposition~\ref{ThSemiprod} stating 
there is a fibered 2-covering $\pi: \FC\times\FC\to U$.
We will denote the lift of $\theta$ to
$\FC\times\FC$ by  $\Theta$.
Integrals $h_1,h_2$ can be also lifted to $\FC\times \FC$.
We consider new integrals $f_1,f_2$ on $\FC\times \FC$
which define the same Liouville
foliation and such that $f_1$ (resp. $f_2$) is a Morse function
on the first (resp. second) factor of $\FC\times \FC$
and is constant on the second (resp. first) factor.
Functions $f_1,f_2$ can be projected onto $U$.
The momentum map $(f_1,f_2)$ differs from $(h_1,h_2)$
by a diffeomorphism $\R^2\to \R^2$.
Consequently, we can prove Proposition~\ref{ThLattice}
for integrals $(f_1,f_2)$ instead of $(h_1,h_2)$
Further $\F$ will denote the momentum map $(f_1,f_2)$.

{\it Step 2. Actions on the direct product.}
Let us define two functions $a_1,a_2$
on
$\FC\times\FC$
as follows:
$a_1(p,q):=1/2\smallint_{L(p)\times\{q\}}\Theta$
and 
$a_2(p,q):=1/2\smallint_{\{p\}\times L(q)}\Theta$
where $(p,q)\in \FC\times\FC$
and $L(x)$ denotes the fiber through $x\in \FC$
on $\FC$.
Recall that all fibers on $\FC$
except for the singular one
are a disjoint union of two circles.

{\it Step 3. Proof of part (a): actions on the semi-direct product.}
Functions $a_1,a_2$ are
constant on the fibers
of $\FC\times\FC$ and thus push forward
to $U=\pi (\FC\times\FC)$.
As functions on $U$, they are 
obtained by integrating $\theta$ along the projections
$\gamma_1,\gamma_2$
of cycles
${L(p)\times\{q\}}$,
${\{p\}\times L(q)}$.
Each of these projections consists
of 2 circles belonging to different Liouville tori
on $U$. Hovewever, the integrals of $\theta$
along the two circles are the same by
Condition~\ref{CondSym}.
Fix a connected component
$\gamma_i'$ of $\gamma_i$,
$i=1,2$.

The problem is that 
the cycles $\gamma_1',\gamma_2'$
may not
constitute a basis on the corresponding Liouville 
torus on $U$. They can generate a group of finite index instead.
Proposition~\ref{ThSemiprod} implies by
elementary geometric arguments that
the cycles
$\gamma_1',\gamma_2'$ are a basis
on the corresponding Liouville torus $T$
if $\F^{-1}(\F(T))$ consists of 2 tori
and generate a subgroup of index 2
if $\F^{-1}(\F(T))$ consists of 4 tori.
\footnote{The latter happens
when
${L(p)\times\{q\}}$ and
${\{p\}\times L(q)}$
both belong to the darker area on $C_2$ on fig.~\ref{figC2flat}.
In this case the involution $(\alpha,\alpha)$
preserves each of the two circles of these fibers.
Each of the 4 corresponding Liouville tori on $\FC\times\FC$
is the product $S^1_a\times S^1_b$
of two circles
that are connected components of
${L(p)\times\{q\}},
{\{p\}\times L(q)}$ (respectively).
The involution $(\alpha,\alpha)$ rotates
both circles by $180^\circ$.
Let $\pi:S^1_a\times S^1_b\to S^1_a\times S^1_b/(\alpha,\alpha)$
be the projection.
Then $\pi(S^1_a),\pi(S^1_b)$ is not a basis
on $\pi(S^1_a\times S^1_b)$,
Instead, $(\pi(S^1_a)\pm\pi(S^1_b))/2$ is a basis.
This is a simple topological fact.
}
In the first case, $a_1,a_2$ are action variables.
In the latter case, cycles $(\gamma_1'\pm\gamma_2')/2$
are a basis on $T$, and the corresponding actions
are $(a_1\pm a_2)/2$. Part (a) is proved.

{\it Step 4. Actions on the plane.}
From now on, we assume that $f_1,f_2$ and $a_1,a_2$ equal $0$
on the $\F$-fiber of point $P$. (Functions $a_1,a_2$ are such for
a well-chosen form $\theta$.)
Functions $a_1,a_2$ depend only on the integrals $f_1,f_2$
and can thus be considered as functions on 
the domain $V=\F(U)\subset \R^2$.
Now we will look at $f_1,f_2$ just as on coordinates
on $V\subset \R^2$.
By definition of $f_1,f_2$ in Step~1, 
the bifurcation diagram is given by
two lines
$\{f_1=0\}\cup\{f_2=0\}$.
The goal of this step is to prove
that
$$
a_i(f_1,f_2)=b_i(f_1,f_2)f_i \ln |f_i|+c_i(f_1,f_2)
$$
where $b_i,c_i$ are smooth functions, 
$c_i(0,0)=0$, $b_i(0,0)\neq 0$.
\footnote{In fact it can be shown
that
$a_1(f_1,f_2)=f_1\ln |f_1|+c_1(f_1,f_2)$,
$a_2(f_1,f_2)=f_2\ln |f_2|+c_2(f_1,f_2)$
for well-chosen integrals $f_i$
(inducing the same Liouville foliation
as $h_i$)
and smooth $c_1,c_2$,
cf. \cite{BS}.
}

First, let us show that
$$
a_i(f_1,f_2)=d_i(f_1,f_2)\ln |d_i(f_1,f_2)|+e_i(f_1,f_2)
$$
where $b_i$ and $d_i$ are smooth and the following
properties hold:
$d_i(0,0)=0$, $d_1(0,f_2)=0$, $d_2(f_1,0)=0$,
$\bd_{f_1}d_1(0,f_2)\neq 0$,
$\bd_{f_2}d_2(f_1,0)\neq 0$.
Indeed, for each fixed value of $f_2$,
consider $a_1(f_1,f_2)$ as function of one variable $f_1$
with parameter $f_2$.
It is just the action function on the 2-dimensional
manifold $C_2\times\{q\}$ for some $q\in C_2$.
It is well-known that $a_1=d_1(f_1,f_2)\ln |d_1(f_1,f_2)|+e_1(f_1,f_2)$
as function of $f_1$. Here $d_1,e_1$ are smooth functions
of $f_1$ with parameter $f_2$. They satisfy the above properties.
It is easy to show that $d_1,e_1$ depend smoothly on $f_2$.
The equality from this paragraph is proved for $a_1$;
$a_2$ is considered analogously.

Now, to prove the initial equality,
observe that
the above properties imply
$d_i=f_i b_i(f_1,f_2)$
for smooth $b_i$ such that $d_i(0,0)\neq 0$.
It suffices to make the substitution
$\ln|d_i|=\ln|f_i|+\ln|b_i|$
and note that 
$\ln|b_i|$ is a smooth function in a neighborhood of $(0,0)$.

{\it Step 5. Proof of part (b).}
Using Step 4, we will show that the map
$\psi:(f_1,f_2)\mapsto (a_1(f_1,f_2),a_2(f_1,f_2))$
is a local homeomorphism at $\F(P)=(0,0)$.
Consider the homeomorphism 
$\phi:(f_1,f_2)\mapsto ((f_1\ln|f_1|)^{-1}$, $(f_2\ln|f_2|)^{-1})$.
It takes functions $a_i$ to
$$
\alpha_i=f_i b_i((f_1\ln|f_1|)^{-1},(f_2\ln|f_2|)^{-1})+c_i((f_1\ln|f_1|)^{-1},(f_2\ln|f_2|)^{-1}).
$$
They are $C^1$-smooth because
$(f_i\ln|f_i|)^{-1}$ are $C^1$-smooth.
Moreover, the differential of $\alpha_1,\alpha_2$ at $(0,0)$
equals
$\mathrm{diag}(b_1(0,0),b_2(0,0))$
since
$\bd_{f_i}(f_i\ln|f_i|)^{-1}(0)=0$.
This differential is non-degenerate,
so $(\alpha_1,\alpha_2)$ is a local homeomorphism.
Then $\psi=(\alpha_1,\alpha_2)\circ\phi$ is also
a local homeomorphism.
Other statements of part (b) are simple.

Finally, part (c) of Proposition~\ref{ThLattice}
is a straightforward corollary of parts (a), (b).~$\blacksquare$

\medskip
{\bf Acknowledgements.}
The author is grateful 
to A.V.~Bolsinov and A.T.~Fomenko for fruitful
discussions and constant support,
and to San
V\~u Ng\d{o}c 
and B.I.~Zhilinskii for encouraging comments.

\smallskip
The work was presented at 
conferences
``Geometry, Dynamics, Integrable systems''
GDIS 2010
(Belgrade, Serbia),
``Finite Dimensional Integrable Systems in Geometry and Mathematical Physics''
FDIS 2011
(Jena, Germany)
and
``Bihamiltonian Geometry and Integrable Systems''
BGIS 2011
(Bedlewo, Poland).
The author is grateful to the organizers of the above conferences
for their hospitality.


\end{document}